\documentclass[aps,prl,reprint,showpacs,floatfix]{revtex4-1}

\usepackage{amsmath}
\usepackage{slashed}
\usepackage{txfonts}
\usepackage{microtype}
\usepackage{graphicx}
\usepackage[breaklinks=true]{hyperref}
\usepackage{color}

\def\sout{\bgroup\markoverwith
{\textcolor{red}{\rule[0.5ex]{2pt}{0.5pt}}}\ULon}
\def\be{\begin{equation}}
\def\ee{\end{equation}}
\def\bes{\begin{equation*}}
\def\ees{\end{equation*}}
\def\bea{\begin{eqnarray}}
\def\eea{\end{eqnarray}}
\def\beas{\begin{eqnarray*}}
\def\eeas{\end{eqnarray*}}
\def\bal#1\eal{\begin{align}#1\end{align}}
\def\bals#1\eals{\begin{align*}#1\end{align*}}
\newcommand{\bra}[1]{\langle #1|}
\newcommand{\ket}[1]{|#1\rangle}
\newcommand{\braket}[2]{\langle #1|#2\rangle}
\newcommand{\bk}[1]{\langle #1\rangle}

\renewcommand{\vec}{\vectorsym}
\newcommand{\del}{\partial}
\renewcommand*{\vec}[1]{\boldsymbol{#1}}

\graphicspath{{figures/}}

\usepackage[normalem]{ulem}

\begin{document}

\title{Anyonic statistics of quantum impurities in two dimensions}  

\author{E. Yakaboylu}
\email{enderalp.yakaboylu@ist.ac.at}
\author{M. Lemeshko}
\email{mikhail.lemeshko@ist.ac.at}

\affiliation{IST Austria (Institute of Science and Technology Austria), Am Campus 1, 3400 Klosterneuburg, Austria}

\date{\today}

\begin{abstract}

We demonstrate that identical impurities immersed in a two-dimensional many-particle bath can be viewed as flux-tube-charged-particle composites described by fractional statistics. In particular, we find that the bath manifests itself as an external magnetic flux tube with respect to the impurities, and hence the time-reversal symmetry is broken for the effective Hamiltonian describing the impurities. The emerging flux tube acts as a statistical gauge field after a certain critical coupling. This critical coupling corresponds to the intersection point between the quasiparticle state and the phonon wing, where the angular momentum is transferred from the impurity to the bath. This amounts to a novel configuration with emerging anyons. The proposed setup paves the way to realizing  anyons using electrons interacting with superfluid helium or lattice phonons, as well as using atomic impurities in ultracold gases.

\end{abstract}

\maketitle

The spin-statistics theorem can be elegantly explained within the unification of quantum mechanics and special relativity under the name of quantum field theory~\cite{Pauli_58,Schwinger_51,burgoyne1958connection,greiner2013field}. There, the so-called microcausality, which guarantees the Lorentz invariance of the $S$ matrix, immediately leads to the right statistics~\cite{Weinberg_64}. Accordingly, particles with integer spins obey Bose-Einstein statistics, whereas half-integer-spin particles are fermions. In a more compact way, when two identical particles are swapped, the wave function of the system is either symmetric or anti-symmetric depending on the particles' spin,~$s$:
\be
\label{spin_stat}
\ket{\psi_1 \psi_2} = (-1)^{2 s} \ket{\psi_2 \psi_1} \, . 
\ee

This is, however,  true only in three spatial dimensions -- in a two-dimensional world the situation is drastically different. As pointed out for the first time by Leinaas and Myrheim~\cite{leinaas1977theory}, in the two-dimensional case,   exchange of two identical particles induces an arbitrary phase, because of the topological basis of two spatial dimensions where the world lines of particles can braid around each other~\cite{leinaas1977theory,goldin1981representations,Wu_84}. At the fundamental level this reflects the fact that spin is not quantized in a $2+1$ dimensional spacetime, as the corresponding little group of the Poincar\'e group is given by SO(2), where there is only one single axis of rotation~\cite{Jackiw_91,barnich2014notes}.

If we consider the spin-statistics theorem~(\ref{spin_stat}) as a general rule, an arbitrary spin value immediately yields the condition $\ket{\psi_1 \psi_2} = e^{i \xi} \, \ket{\psi_2 \psi_1}$ with $\xi \equiv 2 \pi s$. Here the so-called statistical parameter $\xi$ identifies the  statistical  nature of the system: while $\xi = 0 $ for bosons and $\xi = \pi $ for fermions,  in general, $\xi$ can assume any intermediate values. In relative coordinates of two particles, $(r, \varphi)$, this condition can be written as
\be
\label{spin_stat_2}
\psi'(r, \varphi + \pi) = e^{i \xi } \,  \psi'(r, \varphi) \, .
\ee
This implies, however, an unusual boundary condition, $\psi'(r, \varphi + 2 \pi) = e^{2 i \xi } \,  \psi'(r, \varphi)$. Nevertheless, assuming that $\psi'(r, \varphi)$ is an eigenstate of a Hamiltonian $\hat H'$, one can introduce a single-valued wave function, $\psi(r, \varphi) = \exp[-  2 i \xi \varphi / (2\pi)] \psi'(r, \varphi)$, which is governed by the Hamiltonian
\be
\label{anyon_ham}
 e^{-2 i \xi \varphi / (2 \pi) } \, \hat H' \left\lbrace \frac{\del}{\del \varphi}  \right\rbrace\, e^{2 i \xi \varphi / (2 \pi) } = \hat H \left\lbrace \frac{\del}{\del \varphi} + i \frac{2\xi}{2\pi} \right\rbrace \, ,
\ee
where $\lbrace ... \rbrace$ on the left-hand side implies that the $\hat H'$ operator contains a term proportional to $\del/\del\varphi$, and similarly for the right-hand side (units of $\hbar \equiv 1$ are used hereafter). The appearance of the parameter $\xi$ in the Hamiltonian establishes a connection between the spin-statistic theorem and gauge fields~\cite{leinaas1977theory,Wilczek_82}. Namely, the particles that obey the free Schr\"{o}dinger equation are described by fractional statistics in the primed gauge, whereas in the unprimed gauge they turn into bosons interacting with an effective magnetic gauge field. Furthermore, the so-called statistical gauge field, $2\xi /(2\pi) $, implies that orbital angular momentum of two particles in relative coordinates is fractional (and can, in fact, assume any value). In other words, fractionalization of the angular momentum in relative coordinates indicates fractional statistics~\cite{Zhang_2014}.

After all, we do live in an (at least) $3+1$ dimensional spacetime, and spin is quantized. Nevertheless, the role of the statistical gauge field can be substituted by a magnetic gauge field as long as it induces a topological phase. For instance, if we exchange two charged particles inside a constant magnetic field or by enclosing the magnetic flux of a solenoid as in the case of the AB effect~\cite{Aharonov_Bohm_1959}, we naturally obtain a phase factor in the form of Eq.~(\ref{spin_stat_2}), with $\xi$ given by the magnetic flux. Such kind of configurations, however, do not induce a statistical phase. This is because there either the flux depends on the path of the exchange loop or there exists an exchange loop that may not enclose the solenoid at all. A magnetic field can induce a statistical phase if the resulting flux depends only on the winding number of the exchange loop. The latter configuration, on the other hand, can be realized in the following way. If one attaches a flux tube to each of the charged particles and swaps the resulting flux-tube-charged-particle composites, the exchange loop always encloses the flux of one of the composites. Furthermore, it will be the same for all possible exchange loops so that the emerging phase becomes statistical or topological, i.e., independent of the geometry of the exchange loop. Therefore, the flux-tube-charged-particle composites are described by fractional statistics. This is, in fact, exactly what Wilczek considered in order to introduce the concept of the anyon, a particle that obeys \textit{any} statistics~\cite{Wilczek_82,Wilczek_82b}.

Since Wilczek coined the term anyon, there has been a large amount of studies of  both Abelian and non-Abelian anyons. Both of them were predicted to be realized in certain fractional quantum Hall systems~\cite{moore1991nonabelions,Arovas_84,Halperin_84,Nayak_08,Cooper_2015,Lundholm_2016}. In particular, non-Abelian statistics has received a significant amount of attention, as it enables unitary gate operations   necessary for quantum computation~\cite{kitaev2003fault,freedman2003topological,ogburn1999topological,Sarma_05} (also see Ref.~\cite{Nayak_08}, and references therein). Apart from the fractional quantum Hall configurations, emerging Abelian and non-Abelian anyons have also been studied, with
experimental proposals, in several systems~\cite{read2000paired,duan2003controlling,douccot2004discrete,micheli2006toolbox,zhang2007anyonic,yao2007exact,dusuel2008creation} based on the Kitaev model~\cite{kitaev2003fault,kitaev2006anyons}.  

In this paper, we consider two identical impurities immersed in a two-dimensional many-particle bath, and show that the emerging quasiparticle can be seen as a charged particle interacting with a gauge field of a flux tube in relative coordinates. The emerging gauge field is the manifestation of the many-particle bath with respect to the impurity, and acts as a statistical gauge field after a certain critical coupling, leading to fractional statistics for impurities. Our proposal gives a promising opportunity for  experimental observation of anyons in state-of-the-art experiments on two-dimensional materials and quantum liquids.

Let us start by considering two indentical non-interacting impurities trapped in a two-dimensional (2D) bosonic bath, which can be realized with atoms in a 2D Bose-Einstein condensate as well as electrons on superfluid helium films or in 2D polar semiconductors and ionic crystals. At the Fr\"{o}hlich level~\cite{frohlich1954electrons}, the corresponding Hamiltonian is given by:
\bal
\label{two_imp}
 & \hat H_\text{2imp} = \frac{1}{2m} \hat{\vec{P}}_1^2 + \frac{1}{2m} \hat{\vec{P}}_2^2 +  \sum_{\vec{k}} \omega (k) \hat b^\dagger_{\vec{k}} \hat b_{\vec{k}} \\
\nonumber & +  \sum_{\vec{k}} V(k) \left[ \left( e^{-i \vec{k}\cdot \hat{\vec{x}}_1 } +e^{-i \vec{k}\cdot \hat{\vec{x}}_2 } \right) \hat b^\dagger_{\vec{k}} + \text{H.c.} \right]  \, ,
\eal
with  $\sum_{\vec{k}} \equiv \int d^2 k / (2 \pi)^2$. Here $\hat{\vec{P}}_i$ and $\hat{\vec{x}}_i $ are, respectively, the linear momentum operator and the coordinate of each impurity with mass $m$. The third term corresponds to the kinetic energy of the bosons parametrized by the dispersion relation, $\omega(k)$. The bosonic creation and annihilation operators, $\hat b^\dagger_{\vec{k}}$ and $\hat b_{\vec{k}}$, obey the commutation relation $[\hat b_{\vec{k}}, \hat b^\dagger_{\vec{k}'}] = (2 \pi)^2 \delta^{(2)} (\vec{k} - \vec{k}')$.  The second line in Eq.~(\ref{two_imp}) describes the interaction between the impurities and the bosonic bath with the coupling strength $V(k)$, and H.c. stands for the Hermitian conjugate.

First, we introduce relative and center-of-mass coordinates of the impurities: $ \hat{\vec{x}} = \hat{\vec{x}}_2 - \hat{\vec{x}}_1 $, $\hat{\vec{X}} = (\hat{\vec{x}}_2 + \hat{\vec{x}}_1)/2$, which yield the linear momentum operators: $ \hat{\vec{P}}_x  = (\hat{\vec{P}}_2 - \hat{\vec{P}}_1)/2$, $ \hat{\vec{P}}_X  = \hat{\vec{P}}_2 +  \hat{\vec{P}}_1$. In terms of the new coordinates, Eq.~(\ref{two_imp}) can be rewritten as:
\bal
\label{two_imp_2}
 & \hat H_\text{2imp} = \frac{1}{4m} \hat{\vec{P}}_X^2 + \frac{1}{m} \hat{\vec{P}}_x^2  + \sum_{\vec{k}} \omega (k) \hat b^\dagger_{\vec{k}} \hat b_{\vec{k}} \\
\nonumber & + 2 \sum_{\vec{k}} V(k) \cos\left(\vec{k}\cdot \hat{\vec{x}}/2 \right) \left[ e^{-i \vec{k}\cdot \hat{\vec{X}} }  \hat b^\dagger_{\vec{k}} + \text{H.c.} \right] \, .
\eal
The Hamiltonian~(\ref{two_imp_2}) commutes with the total linear momentum of the system, $ \hat{\vec{\Pi}} = \hat{\vec{P}}_X + \sum_{\vec{k}} \vec{k} \, \hat b^\dagger_{\vec{k}} \hat b_{\vec{k}}$. Therefore, if we apply the unitary Lee-Low-Pines transformation~\cite{LLP_53},
\be
\hat T = \exp \left [ - i \hat{\vec{X}} \cdot \sum_{\vec{k}} \vec{k}\,  \hat b^\dagger_{\vec{k}} \hat b_{\vec{k}}  \right] \, ,
\ee
the center-of-mass momentum $\hat{\vec{P}}_X $ becomes a constant of motion in this translated frame. After setting its eigenvalue to zero (which corresponds to the zero total linear momentum in the original frame of Eq.~(\ref{two_imp_2})), the transformed Hamiltonian reads:
\bal
\label{rel_ham}
 & \hat H_\text{rel} \equiv \hat T^{-1}  \hat H_\text{2imp} \hat T = \frac{1}{m} \hat{\vec{P}}_x^2 + \sum_{\vec{k}} \tilde{\omega} (k) \hat b^\dagger_{\vec{k}} \hat b_{\vec{k}}  \\
\nonumber &  + 2 \sum_{\vec{k}} V(k) \cos\left(\vec{k}\cdot \hat{\vec{x}}/2 \right) \left[ \hat b^\dagger_{\vec{k}} + \hat b_{\vec{k}} \right] + \frac{1}{4m}\hat{\Gamma}\, ,
\eal
where $\tilde{\omega} (k) = \omega (k) + k^2/(4 m)$, and $\hat{\Gamma} = \sum_{\vec{k},\vec{k}'} \vec{k} \cdot \vec{k}' \, \hat b^\dagger_{\vec{k}} \hat b^\dagger_{\vec{k}'} \hat b_{\vec{k}} \hat b_{\vec{k}'} $ is the effective phonon-phonon interaction.

Thus, the two-impurity problem reduces to a single-impurity problem in relative coordinates of the two impurities. Next, we decompose the creation and annihilation operators in polar coordinates,
\be
\hat b^\dagger_{\vec{k}} = \sqrt{\frac{2 \pi}{k}} \sum_\mu i^\mu e^{-i \mu \varphi_k} \hat b^\dagger_{k \mu} \, ,
\ee 
such that $ \left[ \hat b_{k \mu}, \hat b^\dagger_{k' \mu'}\right]  = \delta(k - k')\delta_{\mu \mu'} $. Then, the Hamiltonian~(\ref{rel_ham}) can be rewritten as
\bal
\label{rel_ham_2}
 \hat H_\text{rel} & = \frac{1}{m \hat{r}^2} \hat{L}_z^2 + \frac{1}{m} \hat{P}_r^2  + \sum_{k \mu} \tilde{\omega} (k)  \hat b^\dagger_{k \mu} \hat b_{k \mu}  \\
\nonumber &  +  \sum_{k \mu} Y_\mu (k, \hat{r}) \left[  e^{-i \mu \hat \varphi }  \hat b^\dagger_{k \mu} + e^{i \mu \hat \varphi }  \hat b_{k \mu} \right]  + \frac{1}{4m}\hat{\Gamma}' \, ,
\eal
with $\sum_k \equiv \int d k$. Here $\hat L_z \equiv - i \del / \del \varphi $ is the azimuthal angular momentum operator of the two impurities in relative coordinates, and $\hat P_r^2$ is the radial part of $\hat{\vec{P}}_x^2$~\cite{paz2001connection}. Furthermore, $\hat{\Gamma}' = \sum_{k \mu k' \mu'} k k' \,  \hat b^\dagger_{k \mu} \hat b^\dagger_{k' \mu'} \hat b_{k \mu-1} \hat b_{k' \mu'+1} $ is the corresponding effective phonon-phonon interaction in polar coordinates. The impurity-bath coupling strength, on the other hand, is given by:
\be
\label{coupling_in_polar}
Y_\mu (k, \hat{r}) =  \sqrt{k/(2\pi)} \, V(k) J_\mu (k \hat{r} /2) \left[ 1+ (-1)^\mu \right] \, ,
\ee
where we used the Jacobi-Anger expansion, $\exp[i \vec{k} \cdot \vec{x}] =  \sum_\mu i^\mu J_\mu (kr) \exp[i \mu (\varphi - \varphi_k)]$, with $J_\mu (kr)$ being the Bessel function of the first kind. 

We are interested in the properties of the system under particle exchange, which affects only the relative angle $\varphi$ (cf.\ Eq.~(\ref{spin_stat_2})). Accordingly, we assume that the change of the distance between two impurities is very slow compared to its angular motion. In this adiabatic limit we can omit the radial kinetic energy. We can further neglect the phonon-phonon interaction $\Gamma'$, as its expectation value in single bath excitations vanishes. Thereby, the Hamiltonian~(\ref{rel_ham_2}) reduces to
\be
\label{H_phi}
 \hat H_\varphi  = B \hat{L}_z^2  + \sum_{k \mu} \tilde{\omega} (k)  \hat b^\dagger_{k \mu} \hat b_{k \mu}    +  \sum_{k \mu} Y_\mu (k, r) \left[  e^{-i \mu \hat \varphi }  \hat b^\dagger_{k \mu} + e^{i \mu \hat \varphi }  \hat b_{k \mu} \right]  \, ,
\ee
where $B = 1/(mr^2)$. Equation (\ref{H_phi}) is the main Hamiltonian that we are interested in. It describes the relative angular motion of two impurities immersed in a 2D bath, whose interaction with the bath depends on the relative distance $r$. In fact, the Hamiltonian~(\ref{H_phi}) can also be used to describe a system of a quantum planar rotor interacting with a 2D many-particle environment.

It is straightforward to show that the Hamiltonian~(\ref{H_phi}) commutes with the total angular momentum of the impurity-bath system, $ \hat J_z = \hat L_z + \hat \Lambda_z $, where $\hat  \Lambda_z = \sum_{k \mu} \, \mu \, \hat b^\dagger_{k \mu} \hat b_{k \mu}$ is the collective angular momentum operator of the many-body bath, such that $\hat \Lambda_z \, \hat b^\dagger_{k \mu} \ket{0} = \mu \,  \hat b^\dagger_{k \mu} \ket{0}$. Accordingly, under the canonical transformation 
\be
\label{s_trans}
\hat S = \exp(- i \hat \varphi \otimes \hat  \Lambda_z) \, ,
\ee
the total angular momentum reduces solely to the angular momentum of the impurity $ \hat S^{-1} \hat J_z \hat S = \hat L_z $, and hence the angular momentum part of the impurity decouples from the rest of the Hamiltonian in this co-rotating frame. As a result, we can replace $\hat{L}_z$ with its corresponding eigenvalue, $M$, in which case the Hamiltonian~(\ref{H_phi}) reduces to
\be
\label{H_bos}
\hat H_{\text{bos}}  = B \left(M - \hat \Lambda_z \right)^2 + \sum_{k \mu} \tilde{\omega} (k) \hat b^\dagger_{k \mu} \hat b_{k \mu} + \sum_{k \mu} Y_\mu (k,r) \left[ \hat b^\dagger_{k \mu} +  \hat b_{k \mu}\right] \, .
\ee
Thus, the eigenstate of the Hamiltonian~(\ref{H_phi}) can be written as
\be
\label{eigen_state}
\ket{\Psi} = \hat S  \ket{M} \otimes \ket{\text{bos}_n} \, ,
\ee
where the boson state, $\ket{\text{bos}_n}$, is the eigenstate of the Hamiltonian~(\ref{H_bos}) with some quantum number $n$. We would like to emphasize that although Eq.~(\ref{eigen_state}) is reminiscent of the Born-Oppenheimer approximation, the decoupling performed here is exact and represents a unique feature of a 2D quantum impurity problem. We also note that such a  decoupling is exact even for the most general Hamiltonian~(\ref{rel_ham_2}), as the latter also commutes with the total angular momentum $\hat J_z$.

\begin{figure}
  \centering
  \includegraphics[width=\linewidth]{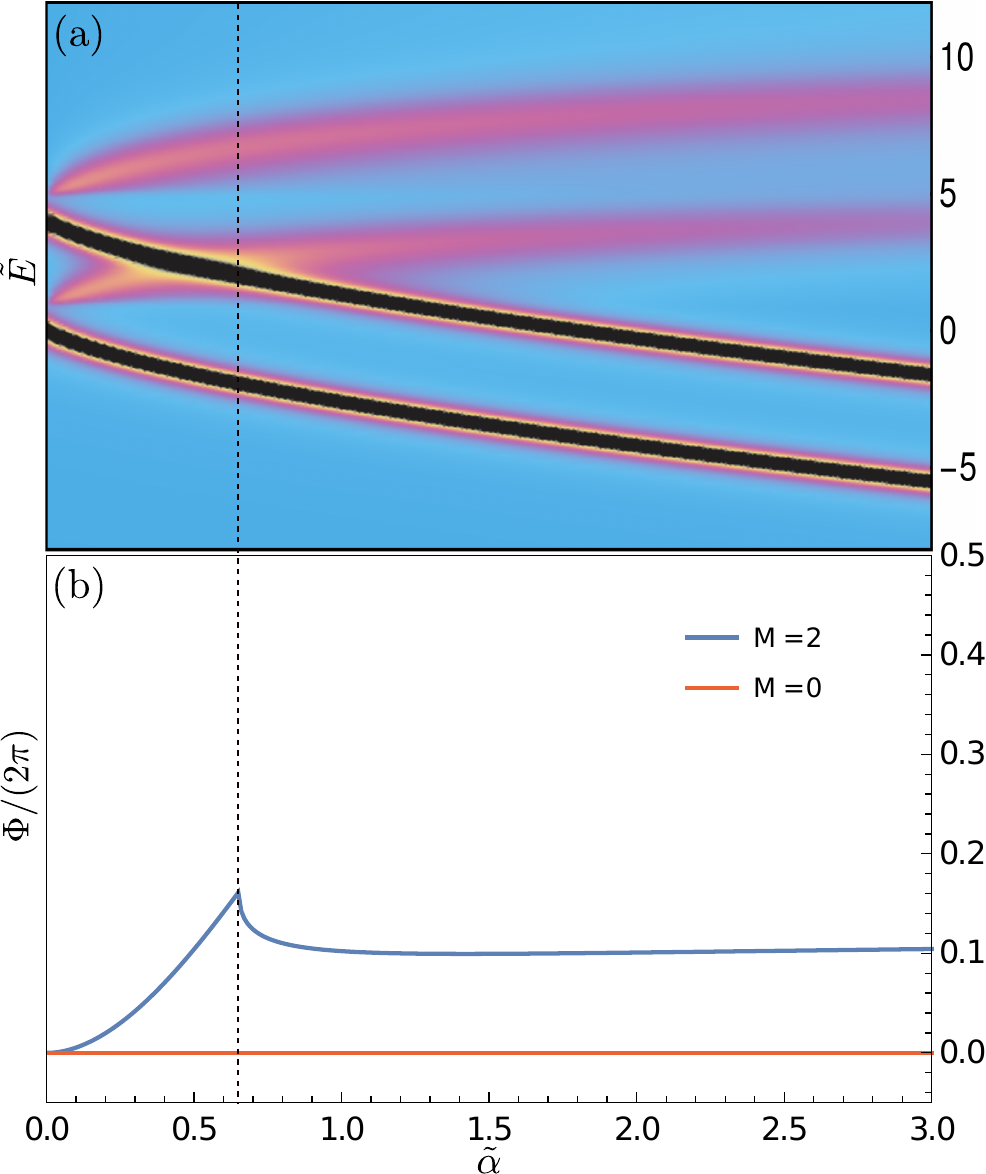}
 \caption{(a) The spectral function of the relative angular motion of two impurities immersed in a 2D bath, $\mathcal{A}_M$, with $\tilde{E} = E/B$, and (b) the magnetic flux $\Phi/(2\pi)$, as a function of the coupling constant $\tilde{\alpha} \equiv \alpha_F/(r B^2)$. The vertical dashed line indicates the critical coupling, after which the magnetic flux becomes constant and behaves as a statistical gauge field. See the text.}
 \label{spec}
\end{figure} 

In the solution~(\ref{eigen_state}), $M$ gives the value of the total angular momentum of the entire system (the impurity plus the many-body environment), where the impurity, in fact, represents two impurities in their relative coordinates. Now, we can ask the following question: What is the angular momentum of two impurities in relative coordinates in the presence of a many-particle bath? The answer to this question can be obtained in the following way. 

In a recent article~\cite{Yakaboylu17}, it was shown that in any impurity problem a many-body environment manifests itself as an external gauge field with respect to the impurity interacting with it. Hence, an impurity problem can be viewed as interaction of a charged particle with this gauge field. Such a formalism  allows one to study geometric and topological properties of impurity problems. Along these lines, in the relative angle $\varphi$, the Hamiltonian~(\ref{H_phi}) can be written as:
\be
\label{H_phi_2}
\hat H_\varphi = - B \del_{\varphi}^2 + \hat H_\text{mb} (\varphi) \, ,
\ee 
with $\hat H_\text{mb} (\varphi)= \sum_{k \mu} \tilde{\omega} (k) \hat b^\dagger_{k \mu} \hat b_{k \mu} + \sum_{k \mu}Y_\mu (k,r) \left[e^{-i \mu \varphi} \hat b^\dagger_{k \mu} \right. + \left. e^{i \mu \varphi} \hat b_{k \mu}\right]$ being the many-body Hamiltonian. It follows from Eq.~(\ref{eigen_state}) that the eigenstate that fulfills the eigenvalue equation
\be
\label{eigen_value_eq}
\hat H_\varphi \ket{\Psi (\varphi) } = E \ket{\Psi (\varphi)} \, ,
\ee
can be decomposed as $\ket{\Psi (\varphi)} \equiv \braket{\varphi}{\Psi} = \chi (\varphi) \ket{\psi_n (\varphi) }$, where $\chi (\varphi) = \braket{\varphi}{M} = \exp(i M \varphi)/ \sqrt{2 \pi}$ and $ \ket{\psi_n (\varphi)} = \hat S(\varphi)  \ket{\text{bos}_n}  $ are the wave function of two impurities in relative coordinates and the many-body bath state, respectively. After we project onto the basis vector $\bra{\psi_n (\varphi) }$, the eigenvalue equation~(\ref{eigen_value_eq}) reduces to the following one for the wave function of the two impurities:
\be
\label{gauge_ham}
 - B \left( \frac{\del}{\del \varphi}  - i A_\varphi \right)^2  \chi (\varphi) = E' \, \chi (\varphi) \, .
\ee
Here $E' = E  -  B \sum_{m \ne n} | \bra{\psi_n (\varphi) } i \del_\varphi \ket{\psi_m (\varphi)}|^2 - \bra{\psi_n (\varphi) } \hat H_\text{mb} \ket{\psi_n (\varphi)} $ is the energy of the impurities interacting with the gauge field
\be
A_\varphi = \bra{\psi_n (\varphi) } i \del_\varphi \ket{\psi_n (\varphi)} = \bra{\text{bos}_n}  \hat \Lambda_z  \ket{\text{bos}_n} = \bk{\hat \Lambda_z} \, ,
\ee
which is the expectation value of the collective angular momentum of the bath. In the Schr\"{o}dinger equation~(\ref{gauge_ham}), the many-body bath manifests itself as an external  magnetic gauge field. Therefore, the corresponding time-reversal symmetry is broken for impurities immersed into the bath. The broken symmetry for impurities, in turn, implies that the time-reversal symmetry for the bosonic Hamiltonian is also broken, which can be seen from Hamiltonian~(\ref{H_bos}), so that the total Hamiltonian~(\ref{H_phi}) remains time-reversal invariant. In other words, as we discuss below, while the angular momentum of the total system is given by the integer~$M$, the impurities' angular momentum in relative coordinates is non-integer.

The magnetic flux, which is given by
\be
\Phi  = \oint A_\varphi \, d \varphi =  2 \pi \, \bk{\hat \Lambda_z} \, ,
\ee
then, can be calculated by finding the eigenstate $\ket{\text{bos}_n}$ from the Hamiltonian~(\ref{H_bos}). Equivalently, it can be calculated in the following way. First, it follows from the Hamiltonian~(\ref{H_bos}) that $\del \hat H_\text{bos} / \del M = 2 B (M - \hat \Lambda_z) $. Then, by using the Hellmann-Feynman theorem, one obtains
\be
\label{flux_hf}
\frac{\Phi}{2\pi} = M - \frac{1}{2 B} \frac{\del E}{\del M} \, ,
\ee 
where $E$ is the corresponding energy eigenvalue of the Hamiltonian~(\ref{H_bos}) or (\ref{H_phi}), and it can be evaluated with the aid of certain variational~\cite{Chevy_06,Lemeshko_2015} as well as renormalization group approaches~\cite{Wilson_75,grusdt2015renormalization},  or diagrammatic Monte Carlo methods~\cite{Prokof_98,Gull_11,anders2011dynamical}. 

\begin{figure}
  \centering
  \includegraphics[width=\linewidth]{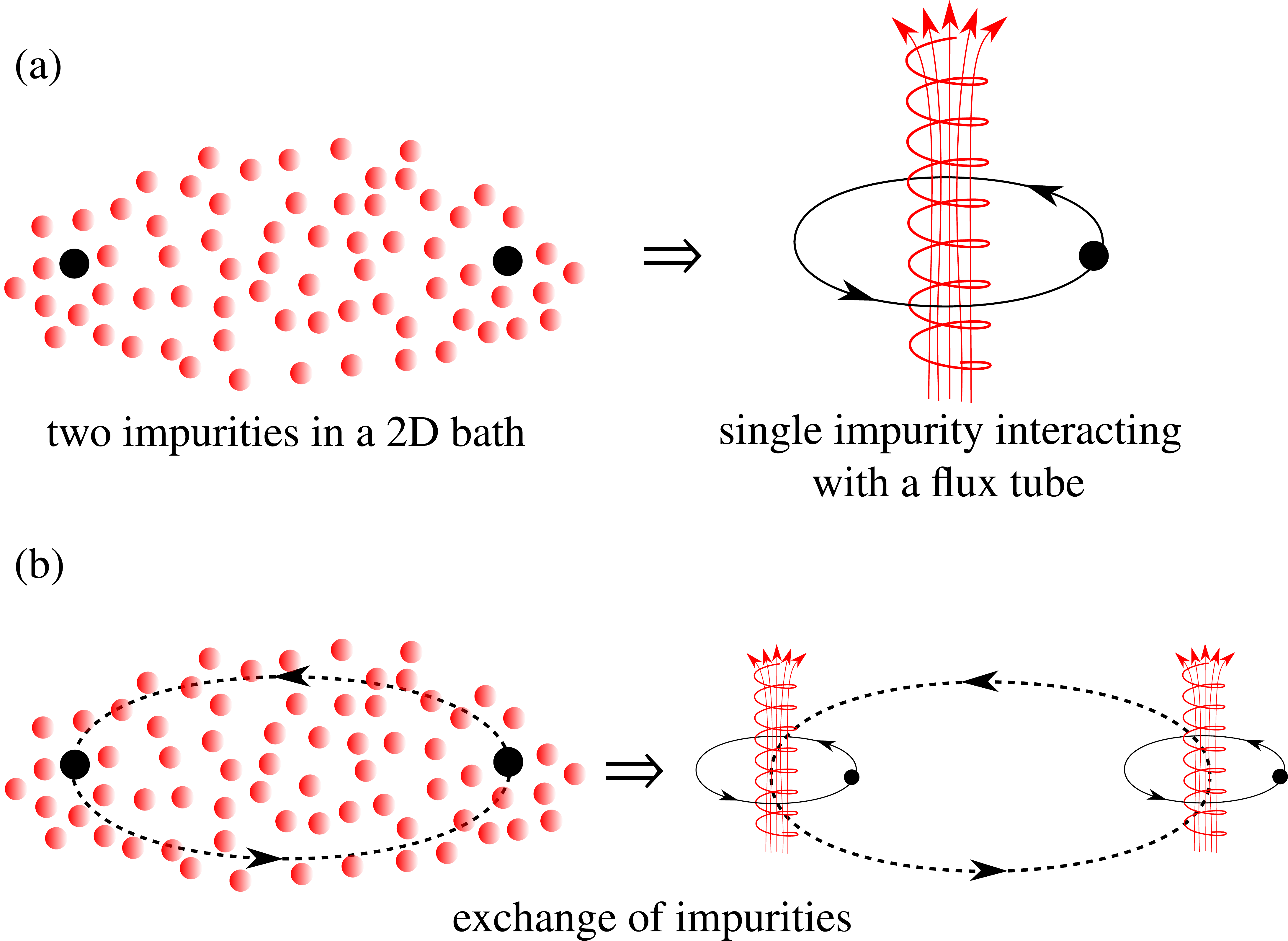}
 \caption{(a)~After the critical coupling, (cf. Fig. 1), two impurities immersed in a two-dimensional bath behave as a charged particle orbiting around a magnetic flux tube in relative coordinates. The latter is the manifestation of the bath with respect to the impurities. Consequently, the angular momentum of the two impurities in relative coordinates becomes fractional. (b)~Inside the bath, the impurities turn into flux-tube-charged-particle composites, and obey anyonic statistics upon exchange. See the text.} 
 \label{flux}
\end{figure} 

Here, we follow the variational approach. For this purpose, we introduce the following variational  ansatz for the Hamiltonian~(\ref{H_phi})~\cite{Chevy_06}:
\be
\label{chevy}
\ket{\Psi_v} = \sqrt{Z} \ket{M} \ket{0} + \sum_{k \mu} \beta_{k \mu} \ket{M-\mu} \hat b^\dagger_{k \mu} \ket{0} \, ,
\ee
which is an eigenstate of the total angular momentum operator $\hat{J}_z$. Here,  $\sqrt{Z}$ and $\beta_{k \mu} $ are the variational parameters with the normalization condition $|Z| + \sum_{k \mu} |\beta_{k \mu}|^2 =1$. Minimization of the functional $\bra{\Psi_v} H -E \ket{\Psi_v}$ with respect to the parameters $\sqrt{Z}^*$ and $\beta_{k \mu}^*$ yields the variational energy, $ E = B M^2 - \Sigma_M (E)$, where 
\be
\Sigma_M (E) = \sum_{k\mu} \frac{Y_\mu (k,r)^2}{B(M-\mu)^2 + \tilde{\omega}(k)-E}
\ee
is the self-energy. The energy is found self-consistently as the solution of the Dyson equation for the following Green's function
\be 
G_M (E) = \frac{1}{B M^2 - \Sigma_M (E) - E} \, ,
\ee
and the entire excitation spectrum of the system is captured by the spectral function $\mathcal{A}_M = \text{Im}\left[G_M (E + i 0^+)\right]$. In Fig.~\ref{spec}~(a), we show the resulting spectral function for the lowest two states, $M=0,2$, by considering a constant dispersion relation, $\omega(k) = \omega_0$, and the coupling strength $ |V(k)| = \sqrt{\sqrt{2}\pi \alpha_F/k}$, with $\alpha_F$ being the Fr\"{o}hlich particle-phonon coupling constant in units of $m=\omega_0=1$~\cite{Wu_86,Peeters_87,devreese2012physics}. Dark sharp peaks in the spectrum correspond to quasiparticle states, and we observe that their energy decreases with increasing coupling $\tilde{\alpha} \equiv \alpha_F/(r B^2)$. The blurred peaks, on the other hand, correspond to the phonon wings.

Next, we calculate the flux from Eq.~(\ref{flux_hf}):
\be
\frac{\Phi}{2\pi} = \frac{\sum_{k\mu} \mu\, Y_\mu (k,r)^2/\left[ B(M-\mu)^2 + \tilde{\omega}(k)-E\right]^2}{1 + \sum_{k\mu} Y_\mu (k,r)^2/\left[ B(M-\mu)^2 + \tilde{\omega}(k)-E\right]^2} \, ,
\ee
which vanishes for the $M=0$ state. In Fig.~\ref{spec}~(b) we show the corresponding flux for the $M=2$ state as a function of the coupling constant $\tilde{\alpha}$. We find that the flux first increases up to a certain value of the coupling constant, but afterwards it saturates, and becomes independent from the coupling parameter and hence the relative distance $r$. This critical coupling corresponds to the point of intersection between the quasiparticle state and the phonon wing, where the angular momentum is transferred from the impurity to the bath.

Since the flux becomes robust after the critical coupling, it can substitute for the role of the statistical gauge field. First, the covariant angular momentum operator of two impurities in relative coordinates reads $ - i \del_\varphi - \Phi/(2 \pi) $, whose eigenvalue is given by $M - \Phi/(2\pi) $. Here, due to the single-valuedness of the wave function $\chi (\varphi + 2 \pi ) = \chi (\varphi )$, the values of $M$ have to be integer. Moreover, if we neglect the spin degree of freedom, $M$ is an even integer due to the fact that in the absence of the bath the spin-statistics theorem requires $\chi (\varphi + \pi ) = \chi (\varphi )$~\cite{Wilczek_82b}. However, because the flux $\Phi$ is a non-integral number, the angular momentum of two impurities in relative coordinates becomes fractional. Thus, in relative coordinates, two impurities confined on a 2D many-body environment effectively behave as a charged particle rotating around a magnetic flux tube. This is the manifestation of the many-body bath with respect to the impurities. $\Phi$ is schematically illustrated in Fig.~\ref{flux}~(a). 

Furthermore, the flux $\Phi$ is the total magnetic flux seen by the two impurities in relative coordinates, and therefore, in analogy to Wilczek's flux-tube-charged-particle composite, $\Phi/2 = \xi$ can be interpreted as the magnetic flux of a flux tube around which each impurity orbits. In fact, if we introduce a gauge,  $ A'_\varphi = A_\varphi - \del_\varphi \eta =  0 $, with $\eta = \Phi \varphi / (2 \pi)$, the corresponding two-impurity wave function in this gauge obeys the free Schr\"{o}dinger equation, and can be written as
\be
{\chi'} (\varphi) = e^{i \Phi \, \varphi / (2\pi)} \chi (\varphi) \, .
\ee
Then, a $\pi$ rotation in relative coordinates, which swaps the two impurities, yields a statistical phase
\be
\label{any_stat}
{\chi'} (\varphi + \pi) = e^{i \xi} {\chi'} (\varphi) \, .
\ee
Thus, in the presence of a bath, each impurity turns into a tightly bound flux-tube-charged-particle composite, which is depicted in Fig.~\ref{flux}~(b). In his original setup, Wilczek introduced the flux tube as a solenoid, around which the particle orbits, and the emerging statistics depends only on the flux of the solenoid, independent from the exchange geometry. In our problem, on the other hand, the flux tube arises as a manifestation of the many-particle bath, and for given impurities and bath, every exchange loop yields the same flux after the critical coupling.

In conclusion, in relative coordinates a two-impurity problem reduces to a problem of a single charged particle orbiting around a magnetic flux tube. We have shown that the emerging gauge field of the magnetic flux behaves as a statistical gauge field after the critical coupling, where the impurity transfers its angular momentum to the bath. Consequently, a $\pi$ rotation in relative coordinates, which corresponds to the exchange of two impurities, induces an additional topological phase on the total wave function. While we presented a formalism for impurities interacting with a bosonic bath, a similar approach can be developed for an environment with Fermi statistics or for Bose-Fermi mixtures. From the experimental point of view, direct measurement of the anyonic statistics corresponds to measuring the fractional value of the angular momentum of the two impurities in relative coordinates. In the context of atomic impurities in ultracold gases, this can potentially be extracted from time-of-flight measurements~\cite{Greiner_01} or  momentum-resolved Bragg scattering~\cite{Stenger_99}.


\begin{acknowledgments}

In Memory of \"{O}zlem Gitmez. We are grateful to A. Deuchert, D. Lundholm, N. Rougerie, and M. Serbyn for valuable discussions. E.Y. acknowledges financial support received from the People Programme (Marie Curie Actions) of the European Union's Seventh Framework Programme (FP7/2007-2013) under REA Grant Agreement No. 291734. M.L. acknowledges support from the Austrian Science Fund (FWF), under Project No. P29902-N27.

\end{acknowledgments}


\bibliography{anyon_ref.bib}

\end{document}